\documentclass[a4paper,11pt]{article}
\usepackage{graphicx}
\usepackage{amsfonts}
\usepackage{amsmath}
\usepackage{hyperref}
\title{How not to secure wireless sensor networks revisited: Even if you say it twice it's still not secure}
\author{Chris J. Mitchell\\Information Security Group, Royal Holloway, University of London\\
\url{www.chrismitchell.net}}
\date{20th November 2020}
\newcommand{\qed}{\nobreak \ifvmode \relax \else
      \ifdim\lastskip<1.5em \hskip-\lastskip
      \hskip1.5em plus0em minus0.5em \fi \nobreak
      \vrule height0.75em width0.5em depth0.25em\fi}
\begin{document}

\maketitle

\begin{abstract}
Two recent papers describe almost exactly the same group key establishment protocol for wireless
sensor networks.  Quite apart from the duplication issue, we show that both protocols are insecure
and should not be used --- a member of a group can successfully impersonate the key generation
centre and persuade any other group member to accept the wrong key value.  This breaks the stated
objectives of the schemes.
\end{abstract}

\section{Introduction} \label{section-intro}

Essentially the same group key establishment protocol aimed at wireless sensor networks has been
presented in two published papers, \cite{Hsu16,Hsu17}, both of which appeared online in early 2016.
The sets of authors of the two papers are slightly different, although three names (Hsu, Harn and
Zhang) appear as authors of both papers.  Thus it is clearly no accident that the same material has
been published twice. To clarify matters, the papers are as follows:
\begin{itemize}
\item \emph{Paper A}, \cite{Hsu16}, by Hsu, Harn, He and Zhang;
\item \emph{Paper B}, \cite{Hsu17}, by Hsu, Harn, Mu, Zhang and Zhu.
\end{itemize}
Paper A \cite{Hsu16} was submitted on January 11th 2016 and accepted for publication on March 2nd
2016. The date of submission of Paper B \cite{Hsu17} is not given but it was published online on
February 2nd 2016.  It thus seems likely that the two papers were submitted and revised at very
similar times.  It is noteworthy that neither paper refers to the other.

The fact that the same material has been published twice is clearly disturbing.  The duplication of
publications is somehow made worse by the fact that, as we discuss below, the scheme described is
obviously insecure. This was pointed out in a March 2018 arXiv document \cite{Mitchell18}, which
refers to Paper B. However, at the time this document was written I was unaware of Paper A,
discovering which has motivated this further note.

The title of this document implicitly refers to another paper, \cite{Mitchell20a}, which describes
attacks against three very closely-related key predistribution schemes, also aimed at wireless
sensor networks.  There is a significant overlap in authorship between the three papers considered
there and the two papers considered here.  It might be that a pattern of behaviour can be
discerned.

The remainder of this short paper is structured as follows.  \S\ref{section-scheme} provides a
brief description of the scheme in Paper A, and the trivial differences from the scheme in Paper B
are also noted.  An attack against this scheme is outlined in \S\ref{section-analysis}. Brief
concluding remarks  are given in \S\ref{section-conclusions}.

\section{The Hsu-Harn-He-Zhang scheme}  \label{section-scheme}

The scheme described in Hsu et al.\ \cite{Hsu16} operates as follows.  The description below is
based closely on reference \cite{Mitchell18}.  The following requirements apply; note that we have
made minor changes to the notation of Hsu et al.\ \cite{Hsu16} for consistency with the March 2018
analysis, \cite{Mitchell18}.
\begin{itemize}
\item The protocol works for a set of users ${\cal U}=\{U_i\}$, all registered with a KGC
    trusted by all users to generate and distribute secret keys.
\item All participants agree on a large integer $m=pq$, where $p$ and $q$ are distinct large
    safe primes.  Hsu et al.\ require all computations to take place in a (sic) finite field
    $\mathbb{K}$ with $m$ elements.  Of course, such a finite field cannot exist, so we assume
    instead that calculations are performed in the commutative ring $\mathbb{Z}_m$\footnote{The
    fact that this is what was intended by the authors becomes clear later in the paper, where
    there are references to calculations being performed `mod $m$'.}; the scheme will work with
    very high probability in such a ring, because the probability of randomly choosing a ring
    element which does not have a multiplicative inverse is vanishingly small given that $p$
    and $q$ are large. Indeed, if this wasn't true then factoring RSA moduli would be easy!
    This is the first of two minor differences from the scheme presented in Paper B,
    \cite{Hsu17}, where it is assumed that calculations take place in a field of $p$ elements
    for a large prime $p$.
\item The participants must also agree a cryptographic hash-function $h$.
\item All participants must agree on the function
    $\mathbf{v}_w:\mathbb{Z}_m\rightarrow(\mathbb{Z}_m)^{w+1}$ defined by:
    \[ \mathbf{v}_w(x) = (1,x,x^2,\ldots,x^w) \]
    (where $w\geq 2$).
\item Every user $U_i$ must have a unique identifier $\mbox{ID}_i$ and a secret key
    $x_i\in\mathbb{Z}_m$ shared with the KGC.
\end{itemize}

Now suppose an \emph{initiator} wishes to arrange for a new secret key to be shared by the members
of a group of users ${\cal U}'$ (${\cal U}'\subseteq{\cal U}$), where ${\cal U}' =
\{U_{z_1},U_{z_2},\ldots,U_{z_t}\}$ for some $t\geq 2$.

The protocol proceeds as follows (where all arithmetic is computed in $\mathbb{Z}_m$).
\begin{enumerate}
\item The initiator sends a request to the KGC with the list of $t$ identifiers $\{\mbox{ID}_i:
    i\in{\cal U}'\}$.
\item The KGC broadcasts the list of identifiers $\{\mbox{ID}_i: i\in{\cal U}'\}$ in response.
\item Each user $U_j\in{\cal U}'$ chooses a fresh random challenge $r_j\in\mathbb{Z}_m$ and
    sends it to the KGC.
\item The KGC performs the following steps.
\begin{enumerate}
\item The KGC randomly chooses a group key $S\in\mathbb{Z}_m$ and a value
    $r_0\in\mathbb{Z}_m$, and assembles the $(t+1)$-tuple
    $\mathbf{r}=(r_0,r_1,r_2,\ldots,r_t)$.
\item For every $i$ ($1\leq i\leq t$) the KGC now computes the inner product
    \[ s_i = ( \mathbf{v}_t(x_{z_i}, \mathbf{r}). \]
    The KGC also computes $u_i=S-s_i$.  Note that this represents the second minor
    difference from the scheme in Paper B, \cite{Hsu17}, where $s_i$ is instead calculated
    as:
    \[ s_i = ( \mathbf{v}_t(x_{z_i}+h_1(x_{z_i}||r_i||r_0)), \mathbf{r}) \]
    where $||$ denotes concatenation of bit strings and $h_1$ is an appropriate
    cryptographic hash function.
\item The KGC now computes the tag \emph{Auth} as
\[
\mbox{\emph{Auth}}=h(S||\mbox{ID}_1||\mbox{ID}_2||\ldots||\mbox{ID}_t||r_0||r_1||r_2||\ldots||
r_t||u_1||u_2||\ldots||u_t) \]

where in assembling the input to $h$, elements of $\mathbb{Z}_m$ are converted to bit
    strings using an agreed representation.
\item Finally, the KGC broadcasts
    \[ \mbox{\emph{Auth}}, r_0, (u_1,u_2,\ldots,u_t) \]
to all members of the group ${\cal U}'$.
\end{enumerate}
\item On receiving the broadcast, user $U_{z_i}\in{\cal U}'$ ($1\leq i\leq t$):
\begin{enumerate}
\item computes
    \[ s_i = ( \mathbf{v}_t(x_{z_i}, \mathbf{r})) \]
using its secret key $x_{z_i}$, the random challenges $r_i$ ($1\leq i\leq t$) sent earlier
in the protocol, and the broadcast value $r_0$;
\item computes the group key as $S=u_i+s_i$; and finally
\item verifies \emph{Auth} by recomputing it using the newly computed group key and the
    values received in the protocol.
\end{enumerate}
\end{enumerate}

\section{Analysis}  \label{section-analysis}

The analysis of the protocol precisely follows the analysis in the 2018 note \cite{Mitchell18}. We
suppose a `victim user' $U_v$ is a member of a group of $t$ users for which a new key is requested,
and that one of the users, $U_x$ say, in the group ${\cal U}'$ is malicious.  We also assume that
$U_x$ can control the channel between the KGC and the victim user $U_v$ so that $U_x$ can modify
what $U_v$ receives in the final KGC broadcast in step 4d.  As we show below, $U_x$ is able to make
$U_v$ accept a key $S^*$ of $U_x$'s choice.

We suppose that the protocol proceeds as described in section~\ref{section-scheme}, where $U_v,
U_x\in{\cal U}'$.  In step 4d, $U_x$ prevents the broadcast from the KGC reaching $U_v$.  Because
$U_x$ is a valid member of ${\cal U}'$, $U_x$ can calculate the secret key $S$ generated and
distributed by the KGC\@. $U_x$ now chooses a different secret key $S^*\in\mathbb{Z}_m$, and
computes
\[ u^*_v= u_v-S+S^* \]
and
\[
\mbox{\emph{Auth}}^*=h_2(S^*||\mbox{ID}_1\ldots||\mbox{ID}_t||r_0||r_1||\ldots||
r_t||u_1||\ldots||u_{v-1}||u^*_v||u_{v+1}||\ldots||u_t). \]

That is, \emph{Auth}$^*$ is computed using the same inputs as \emph{Auth} except that $S$ and $u_v$
are switched to $S^*$ and $u^*_v$. $U_x$ now sends a modified version of the KGC's broadcast to
$U_v$, where \emph{Auth} and $u_v$ are replaced by \emph{Auth}$^*$ and $u^*_v$.  It is
straightforward to see that victim user $U_v$ will compute the secret key as $S^*$, and the tag
\emph{Auth}$^*$ will verify.  The attack is complete.

\section{Concluding remarks}  \label{section-conclusions}

Apart from the double publication issue, it is difficult to go much beyond the conclusions of the
2018 note \cite{Mitchell18}.  Neither Paper A not Paper B provide a rigorous security proof using
state of the art `provable security' techniques, nor do they give a formal security model.  This is
despite the existence of well-established security models within which the security properties of
group key establishment protocols can be established (see, for example, \S 2.7.1 of Boyd et al.\
\cite{Boyd20}). This certainly helps to explain why fundamental flaws exist.

Indeed, the following observation of Liu et al.\ \cite{Liu17} regarding a number of previously
proposed but flawed group key establishment protocols is highly relevant.  `The security proof for
each vulnerable GKD protocol only relies on incomplete or informal arguments.  It can be expected
that they would suffer from attacks'.  We conclude that, although it might be tempting to try to
repair the protocol to address the issues identified, unless a version can be devised with an
accompanying security proof (which may well not be possible without significantly increasing the
complexity) this would be foolhardy since there is a strong chance that flaws will remain.

The American philosopher and psychologist William James (1842-1910) reputedly said `There's nothing
so absurd that if you repeat it often enough, people will believe it'\footnote{See, for example,
\url{http://libertytree.ca/quotes/William.James.Quote.7EE1}}. Clearly in this case twice is not
enough times!


\begin{thebibliography}{1}

\bibitem{Boyd20} C.~Boyd, A.~Mathuria, and D.~Stebila, \emph{Protocols for authentication and
  key establishment}, 2nd ed., Information Security and Cryptography, Springer,
  2020.

\bibitem{Hsu16} C.-F. Hsu, L.~{Harn}, T.~He, and M.-Y. Zhang, \emph{Efficient group key
  transfer protocol for {WSNs}}, IEEE Sensors Journal \textbf{16} (2016),
  no.~11, 4515--4520.

\bibitem{Hsu17} C.-F. Hsu, L.~Harn, Y.~Mu, M.~Zhang, and X.~Zhu, \emph{Computation-efficient
  key establishment in wireless group communications}, Wireless Networks
  \textbf{23} (2017), 289--297.

\bibitem{Liu17} J.~Liu, Y.~Wu, X.~Liu, Y.~Zhang, G.~Xue, W.~Zhou, and S.~Yao, \emph{On the
  (in)security of recent group key establishment protocols}, The Computer
  Journal \textbf{60} (2017), 507--526.

\bibitem{Mitchell18} C.~J. Mitchell, \emph{The {Hsu-Harn-Mu-Zhang-Zhu} group key establishment
  protocol is insecure}, arXiv:1803.05365 [cs.CY],
  \url{http://arxiv.org/abs/1803.05365}, March 2018.

\bibitem{Mitchell20a} \bysame, \emph{How not to secure wireless sensor networks: {A} plethora of
  insecure poynomial-based key pre-distribution schemes}, arXiv:2004.05597
  [cs.CY], \url{http://arxiv.org/abs/2004.05597}, April 2020.

\end{thebibliography}
\providecommand{\bysame}{\leavevmode\hbox to3em{\hrulefill}\thinspace}
\providecommand{\MR}{\relax\ifhmode\unskip\space\fi MR }
\providecommand{\MRhref}[2]{%
  \href{http://www.ams.org/mathscinet-getitem?mr=#1}{#2}
} \providecommand{\href}[2]{#2}

\end{document}